\documentclass[prc,showpacs,floatfix]{revtex4}
\usepackage{bm}
\usepackage{graphicx}
\usepackage{amsmath}
\usepackage{bbold}

\newcommand{\beq}{\begin{equation}}
\newcommand{\eeq}{\end{equation}}
\newcommand{\beqa}{\begin{eqnarray}}
\newcommand{\eeqa}{\end{eqnarray}}

\begin{document}

\title{Photon Scattering with the Lorentz Integral Transform Method
\hfill{\small {\bf MKPH-T-11-14}}\\
}

\author{Giulia Bampa$^1$,
       Winfried Leidemann$^{1,2}$, and 
       Hartmuth Arenh\"ovel$^3$}
\affiliation{
  $^{1}$Dipartimento di Fisica, Universit\`a di Trento,
  I-38100 Trento, Italy\\
  $^{2}$Istituto Nazionale di Fisica Nucleare, Gruppo Collegato
   di Trento, Italy\\
  $^3$Institut f\"{u}r Kernphysik, Johannes Gutenberg-Universit\"{a}t, 
   D-55099 Mainz, Germany}

\date{\today}
\begin{abstract}
The application of the Lorentz integral transform (LIT) method to
photon scattering  off nuclei is presented in general. As an example,
elastic photon scattering off the deuteron in the unretarded dipole
approximation is considered using the LIT method. The inversion of the
integral transform is discussed in detail paying particular attention to the high-energy
contributions in the resonance term. The obtained E1-polarizabilities
are compared to results from the literature. The 
corresponding theoretical cross section is confronted with
experimental results confirming, 
as already known from previous studies, that the E1-contribution is
the most important one at lower energies.  
\end{abstract}

\maketitle

\section{Introduction}
The study of electromagnetic reactions in photoabsorption and photon
scattering on nuclei is an excellent tool to investigate
nuclear structure. In addition, it can also lead to valuable
insights into the properties of the nuclear constituents, the
nucleons, like for example, electric and magnetic polarizabilities. In
this context photon scattering experiments are a particularly
interesting source of information on off-shell properties. On the other hand
genuine microscopic calculations of photon scattering cross sections
are rather complicated 
since the complete nuclear excitation spectrum has to be taken into 
account. Thus it hardly comes as a surprise that in the past such
theoretical efforts were mainly concentrated on the two-nucleon
system. A first realistic calculation of deuteron photon scattering
has been carried out in~\cite{WeA83}. In the last decade quite a few
theoretical investigations have been
performed~\cite{Lev95,KaM99,LeL01,HiG05,HiG10} 
among them also calculations based on chiral effective field 
theory.   

Considering a more complex $A$-nucleon system, as mentioned, one needs
to have under control the corresponding $A$-body continuum. Today this
could in principle be realized for the three-nucleon system, but
many-body calculations of photon scattering cross sections for
systems with $A>3$ are presently out of reach. On the other hand,
there is a particular interest in $^6$Li photon scattering  recently,
and data have already 
been taken at HIGS~\cite{Fel11}. Fortunately, the problem of
calculating the correct many-body continuum wave function can be
avoided by application of the Lorentz integral transform (LIT)
method~\cite{ELO94,EfL07}. In fact, the LIT approach reduces a
scattering 
state problem to a simpler bound-state like problem, which then can be
solved with techniques that usually are applied for bound states. This
leads to an enormous reduction of the complexity of the calculation,
e.g., LIT calculations of the total photoabsorption cross sections
have even been made for the six- and seven-body nuclei taking into
account all possible break-up channels and with full consideration of
the final state interaction~\cite{BaB02,BaA07}. In view of the fact,
that till now the LIT method has not been applied to photon
scattering, we want to demonstrate with the present work the usefulness
of this method for this process choosing as a test case the
deuteron. That has the advantage that first of all, the calculation is
quite simple and furthermore allows the comparison with a conventional
approach. 

In section II, first we briefly review the formal theory of photon
scattering, the low energy limit of the scattering amplitude and the
concept of generalized polarizabilities as basic quantities. Then we
develop the general formalism of how these polarizabilities can be
calculated by the LIT method. The application of this method on the
deuteron as a test case is described in section III. For demonstrating
the method it suffices to limit the explicit calculation to the low
energy regime where the E1 contribution dominates. The results are
presented and discussed in section IV where we also give a summary and
an outlook.  

\section{Formal developments}
We start the formal part with a short resumee of the salient features
of photon scattering off a bound many-body system (for a more detailed
review see e.g.~\cite{Are86}), i.e.\ we consider the process 
\beq
\gamma(\vec k) + N_i(\vec P_i \,) \longrightarrow \gamma(\vec
k^{\,\prime}) + N_f(\vec P_f) \,, 
\eeq
for an incoming photon with
momentum $\vec k$ and polarization $\vec e_{\lambda}$ and an outgoing
photon with momentum $\vec k^{\,\prime}$ and polarization $\vec
e_{\lambda'}^{\,\prime}$ while the system makes a transition from an
initial state with total momentum $\vec P_i$ and intrinsic state
$|i\rangle$ to a final state with total momentum $\vec P_f$ with
intrinsic state state $|f\rangle$. 

\subsection{The photon scattering amplitude}

\begin{figure}[ht]
\centerline{\includegraphics[width=.6\columnwidth]{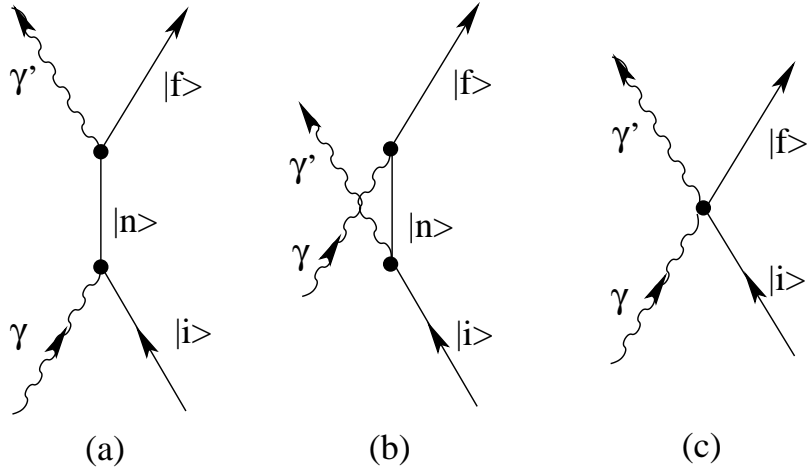}} 
\caption{Diagrammatic representation of the resonance (direct (a) and
  crossed (b))  and the two-photon amplitude (c) for photon scattering.} 
\label{Fig-scattering-diagram} 
\end{figure}

In view of the weakness of the electromagnetic
interaction one can apply perturbation methods. In the lowest,
i.e.~second order in the e.m.~coupling, the scattering amplitude is
given by two terms, the contact or two photon amplitude (TPA) $B
_{\lambda'\lambda}(\vec k^{\,\prime},\vec k)$ and the 
resonance amplitude (RA) $R _{\lambda'\lambda}(\vec k^{\,\prime},\vec
k)$. A graphical illustration is shown in 
Fig.~\ref{Fig-scattering-diagram}. 

Accordingly, the total scattering amplitude is the sum of these two
contributions 
\beq
T^{fi}_{\lambda'\lambda}(\vec k^{\,\prime},\vec k)=
B^{fi}_{\lambda'\lambda}(\vec k^{\,\prime},\vec k)
+R^{fi}_{\lambda'\lambda}(\vec k^{\,\prime},\vec k)\,,\label{scattering-amplitude}
\eeq
where the two-photon amplitude as depicted in diagram (c) of
Fig.~\ref{Fig-scattering-diagram} has the form 
\beq
B^{fi}_{\lambda'\lambda}(\vec k',\vec k)=-\langle f|\int d^3x d^3y
e^{i\vec k^{\,\prime}\cdot\vec x}e^{-i\vec k\cdot\vec y}
\vec e_{\lambda'}^{\,\prime *}\cdot \stackrel{\leftrightarrow}{B}(\vec
  x,\vec y)\cdot \vec e_{\lambda}|i\rangle\,,\label{TP-amplitude}
\eeq
and the resonance amplitude (RA) (diagrams (a) and (b) of
Fig.~\ref{Fig-scattering-diagram})  is given by
\beqa
R^{fi}_{\lambda'\lambda}(\vec k^{\,\prime},\vec k)=
\langle f|&\Big[&\vec e_{\lambda'}^{\,\prime *}\cdot \vec 
  J(-\vec k^{\,\prime},2\vec P_f+\vec k^{\,\prime}) \,G(k+i\epsilon)\,
\vec e_{\lambda}\cdot \vec 
  J(\vec k,2\vec P_i+\vec k)\nonumber\\ && 
+\vec e_{\lambda}\cdot \vec 
  J(\vec k,2\vec P_f-\vec k) \,G(-k'+i\epsilon)\,
\vec e_{\lambda'}^{\,\prime *}\cdot \vec 
  J(-\vec k^{\,\prime},2\vec P_i-\vec k^{\,\prime}) \Big]|i\rangle 
\,,\label{R-amplitude}
\eeqa
with the intermediate propagator
\beq
G(z)=(H-E_i-z)^{-1}\,.
\eeq
 In these expressions, the c.m.~motion has been separated
applying translation invariance. Thus the initial and final states
refer to the intrinsic motion of the many-particle system only. In
principle, the Hamiltonian $H$ and the energy $E_i$ contain
contributions from the
c.m. motion. Furthermore, the splitting in a resonance and a
two-photon amplitude is gauge dependent. This gauge dependence is
reflected in the gauge conditions sketched briefly below. 

The cartesian tensor operator $\stackrel{\leftrightarrow}{B}$ of
rank 2 in
Eq.~(\ref{TP-amplitude}) represents the second order term of
the e.m.~interaction with the system under consideration, and the
current operator in Eq.~(\ref{R-amplitude})
\beq
\vec J(\vec k,\vec P)=\vec j(\vec k)+\frac{\vec P}{2AM}\,\rho(\vec k)\label{current}
\eeq
acts on the intrinsic variables of the system only. It consists of the
intrinsic current $\vec j(\vec k)$ plus a term taking into account the
convection current of the separated c.m.~motion. 
$M$ denotes the nucleon mass and $A$ the mass
number of the nucleus. The intrinsic charge and current operators
consist of a kinetic or one-body and a meson exchange part
\beqa
\rho(\vec k\,)&=&\rho_{[1]}(\vec k\,)+\rho_{[2]}(\vec k\,)\,,\\
\vec j(\vec k\,)&=&\vec j_{[1]}(\vec k\,)+\vec j_{[2]}(\vec k\,)\,,
\eeqa
with
\beqa
\rho_{[1]}(\vec k)&=&\sum_le_l\,e^{-i\vec k\cdot \vec r_l}\,,\\
\vec j_{[1]}(\vec k\,)&=&\frac{1}{2M}\sum_l \Big(e_l \{\vec p_l,e^{-i\vec
  k\cdot \vec r_l}\} + \mu_l \vec \sigma_l\times\vec k\,e^{-i\vec
  k\cdot \vec r_l}\Big)\,.
\eeqa
Here, $e_l$ and $\mu_l$ denote charge and magnetic moment of the
$l$-th particle and $\vec p_l$ and $\vec \sigma_l$ its internal
momentum and spin operator. 
The expressions for the corresponding exchange operators depend on the
interaction model. At least in the nonrelativistic limit, the exchange
contribution to the charge density vanishes (Siegert's hypothesis). 
Furthermore, also the TPA consists of a kinetic one-body contribution
and a two-body exchange amplitude 
\beq
\stackrel{\leftrightarrow}{B}(\vec k^{\,\prime},\vec
k\,)=\stackrel{\leftrightarrow}{B}\! _{[1]}(\vec k^{\,\prime},\vec
k\,)+\stackrel{\leftrightarrow}{B}\! _{[2]}(\vec k^{\,\prime},\vec k\,) \,,
\eeq
where the kinetic one-body operator is given by
\beq
\stackrel{\leftrightarrow}{B}\! _{[1]}(\vec k^{\,\prime},\vec
k\,)= -\frac{1}{M}\sum_l e_l^2 \,e^{-i(\vec k-\vec k^{\,\prime})
\cdot \vec r_l}\,,
\eeq
which is the sum of the individual Thomson amplitudes.

Gauge invariance of the electromagnetic interaction leads to gauge
conditions for the various e.m.~operators according to
\beqa
\vec k\cdot\vec j(\vec k\,)&=&[H,\rho(\vec k\,)]\,,\\
\vec k^{\,\prime}\cdot\stackrel{\leftrightarrow}{B}(\vec
k^{\,\prime},\vec k\,)&=& [\rho(-\vec k^{\,\prime}\,),\vec j(\vec k\,)]\,,
\eeqa
where $H=T+V$ denotes the intrinsic Hamiltonian of the nuclear system
with $T$ as kinetic energy and the interaction potential $V$. 
Separating the one-body and exchange contributions, one finds
\beqa
\vec k\cdot\vec j_{\,[1]}(\vec k\,)&=&[T,\rho_{\,[1]}(\vec k\,)]\,,\\
\vec k\cdot\vec j_{\,[2]}(\vec k\,)&=&[V,\rho_{\,[1]}(\vec
k\,)]+[T,\rho_{\,[2]}(\vec k\,)]\,,\\ 
\vec k^{\,\prime}\cdot\stackrel{\leftrightarrow}{B}\! _{[1]}(\vec
k^{\,\prime},\vec k\,)&=& [\rho_{\,[1]}(-\vec k^{\,\prime}\,),\vec j_{\,[1]}(\vec
k\,)]\,,\\ 
\vec k^{\,\prime}\cdot\stackrel{\leftrightarrow}{B}\! _{[2]}(\vec
k^{\,\prime},\vec k\,)&=& [\rho_{\,[1]}(-\vec k^{\,\prime}\,),\vec j_{\,[2]}(\vec
k\,)]+[\rho_{\,[2]}(-\vec k^{\,\prime}\,),\vec j_{\,[1]}(\vec
k\,)]\,. 
\eeqa
One important consequence are the low energy limits~\cite{Fri95,ArW86}
\beqa
 \vec j(0)&=&[H,\vec D]\,,\\
B^{ii}_{[1],\lambda'\lambda}(0,0)&=&
-\vec e_{\lambda'}^{\,\prime *}\cdot \vec
e_{\lambda}\frac{Ze^2}{M}\,,\\
B^{ii}_{[2],\lambda'\lambda}(0,0)&=&
-\langle i|[\vec e_{\lambda'}^{\,\prime *}\cdot \vec D,[V,\vec
e_{\lambda}\cdot\vec D]]|i\rangle\,,\\
R^{ii}_{\lambda'\lambda}(0,0)&=&
\vec e_{\lambda'}^{\,\prime *}\cdot \vec
e_{\lambda}\frac{NZe^2}{AM}-B^{ii}_{[2],\lambda'\lambda}(0,0)\,,
\eeqa
resulting in the low energy limit for the total scattering amplitude
\beqa
T^{ii}_{\lambda'\lambda}(0,0)&=&
-\vec e_{\lambda'}^{\,\prime *}\cdot \vec
e_{\lambda}\frac{(Ze)^2}{AM}\,,
\label{Thomson}
\eeqa
which is the classical Thomson limit.

\subsection{Generalized nuclear polarizabilities}

The expansion of the scattering amplitude with respect to the total
angular momentum transferred to the nucleus in the scattering process
leads to the concept of generalized polarizabilities. These
polarizabilities allow in a 
convenient manner to separate geometrical aspects related to the
angular momentum properties and dynamical effects given by the strength of
the various polarizabilities. To this end, one starts from the
multipole expansion of the plane wave (see, e.g.~\cite{Ros57})
\beq
\vec e_\lambda\, e^{i\vec k \cdot \vec r}=-\sqrt{2\pi}\sum_{LM}\hat LD^L_{M\lambda}(R)
\sum_{\nu=0,1} \lambda^\nu \vec A^L_M(M^\nu;k)\,,\label{multipolefield}
\eeq 
with standard electric and magnetic multipole fields $\vec A^L_M(M^\nu;k)$,
where $\nu$ indicates the type of multipole field ($M^0=E$ (electric)
and $M^1=M$ (magnetic)). Then one expands the current operator in terms of
electric ($M^{0,L}= E^L$) and magnetic ($M^{1,L}= M^L$) multipole operators 
\beq
\vec e_{\lambda}\cdot\vec j(\vec k)= -\sqrt{2\pi}\sum_{LM}\hat L
D^L_{M\lambda}(R) \sum_{\nu=0,1}  \lambda^\nu M^{\nu,L}_M 
\,,\label{multipolemoments}
\eeq 
where $R$ denotes a rotation which carries the quantization axis into
the direction of $\vec k$, and $D^L_{M\lambda}$ denotes the corresponding
rotation matrix~\cite{Ros57}. A similar expansion holds for the
two-photon operator. 

\begin{figure}[ht]
\centerline{\includegraphics[width=.5\columnwidth]{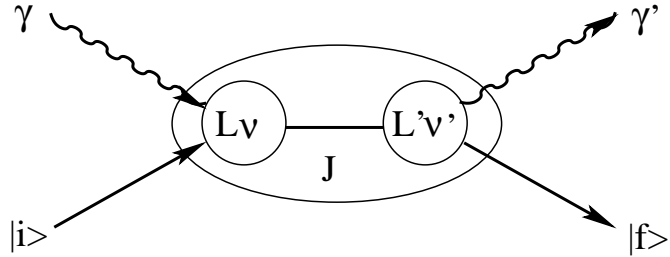}}
\caption{Graphical representation of the contribution of the direct
  term of the resonance amplitude to the generalized polarizability
  $P_{if,J}(M^{\nu'}L',M^{\nu} L,k',k)$ with consecutive angular momentum
  transfers $L$ and $L'$ by multipoles of type $\nu$ and $\nu'$,
  respectively, coupled to total angular momentum transfer $J$.} 
\label{Fig-polarizability} 
\end{figure}

The electric and magnetic multipole
fields of order $L$ of the incoming photon transfer an angular
momentum $L$ according to the strengths of the corresponding nuclear
transition multipole moments. Similarly, the scattered photon
transfers angular momentum $L'$. These consecutive momentum transfers
can further be classified according to the total momentum transfer $J$
to the nucleus with $|L-L'|\le J\le L+L'$. The corresponding strength
is given by the polarizability
\beq
P_{if,J}^{L'L\lambda'\lambda}(k',k)=\sum_{\nu'\nu=0,1}
\lambda^{\prime\,\nu'}\lambda^\nu 
P_{if,J}(M^{\nu'}L',M^{\nu} L,k',k)\,,
\eeq 
where $\nu$ classifies the type of multipole transition as already
mentioned. A graphical visualization of 
the generalized polarizability is shown in Fig.~\ref{Fig-polarizability}.

Then the expansion of the total scattering
amplitude in terms of these polarizabilities reads
\beqa
T^{fi}_{\lambda'\lambda}(\vec k^{\,\prime},\vec k)=
(-)^{1+\lambda'+I_f-M_i}&\sum_{L',M',L,M,J}&(-)^{L+L'}(2J+1)
\left(\begin{matrix}
I_f&J& I_i \cr -M_f &m&M_i \cr
\end{matrix}\right)
\left(\begin{matrix}
L&L'& J \cr M &M'&-m \cr
\end{matrix}\right)\nonumber\\&&
\times P_{if,J}^{L'L\lambda'\lambda}(k',k)D^L_{M,\lambda}(R)
D^{L'}_{M',-\lambda'}(R')\,,
\eeqa
where $(I_i,M_i)$ and $(I_f,M_f)$ refer to the angular momenta and
their projections on the quantization axis of the initial and final states,
respectively. Furthermore, $R$ and $R'$ describe the rotations which
carry the quantization axis into the directions of the photon momenta
$\vec k$ and $\vec k^{\,\prime}$, respectively, and
$D^L_{M,\lambda}(R)$ and $D^{L'}_{M',-\lambda'}(R')$ the corresponding
rotation matrices in the convention of Rose~\cite{Ros57}. 

As for the scattering amplitude, the polarizabilities can be separated
in a TPA  and a resonance contribution
\beq
P_{if,J}(M^{\nu'}L',M^\nu L,k',k)=P_{if,J}^{TPA}(M^{\nu'}L',M^\nu L,k',k)
+P_{if,J}^{\rm res}(M^{\nu'}L',M^\nu L,k',k)\,,
\eeq 
where for the resonance amplitude one has
\beqa\label{compact-res}
P_{if,J}^{\rm res}(M^{\nu'}L',M^{\nu}L,k',k)&=&2\pi(-)^{L+J}\frac{\hat L\hat
  L'}{\hat J}\\
&&\times\langle I_f E_f||\, \Big(\Big[M^{\nu',L'}(k')G(k+i\varepsilon)
M^{\nu, L}(k)\Big]^{J}
+\Big[M^{\nu, L}(k)G(-k'+i\varepsilon)M^{\nu', L'}(k')\Big]^{J}
\Big)||I_i E_i\rangle\,.\nonumber
\eeqa
Here the symbol ``$[\dots]^{J}$'' means that the two multipole
operators are coupled to a spherical tensor of rank $J$. Furthermore,
we have neglected the small c.m. current contribution of
Eq.~(\ref{current}).
The two-photon contribution to the polarizability is given by
\beqa
P_{if,J}^{TPA}(M^{\nu',L'},M^{\nu,L},k',k)&=&2\pi(-)^{L+J+1}\frac{\hat L\hat
  L'}{\hat J}\langle I_f E_f||\int d^3x\, d^3y\, \Big[\vec A^{L'}(
M^{\nu'};k,\vec x)\cdot \stackrel{\leftrightarrow}{B}(\vec
  x,\vec y)\cdot \vec A^L(M^\nu;k,\vec y)\Big]^{J}||
I_i E_i\rangle\,.\nonumber\\
\eeqa
The evaluation of the TPA contribution to the polarizabilities is
straightforward once the TPA operator $\stackrel{\leftrightarrow}{B}(\vec
x,\vec y)$ is given. 

For the resonance contribution, one finds by evaluating the reduced
matrix element in standard fashion (see e.g.~\cite{Edm57})
\beqa
P_{if,J}^{\rm res}(M^{\nu'}L',M^\nu L,k',k)&=&2\pi(-)^{L+I_f+I_i}\hat L\hat L'
\nonumber\\&& 
\times \sum\hspace{-.5cm}\int_{E_n, I_n}\left[
\left\{\begin{matrix}
L&L'& J \cr I_f &I_i& I_n \cr 
\end{matrix}\right\}\,
\frac{\langle I_f E_f||M^{\nu',L'}(k')||I_n E_n\rangle
\langle I_n E_n||M^{\nu,L}(k)||I_i E_i\rangle}{E_n-E_i-k-i\varepsilon}\right.
\nonumber\\&&\left.
+(-)^{L+L'+J}
\left\{\begin{matrix}
L'&L& J \cr I_f &I_i& I_n \cr 
\end{matrix}\right\}\,
\frac{\langle I_f E_f||M^{\nu,L}(k)||I_n\rangle
\langle I_n E_n||M^{\nu',L'}(k')||I_i E_i\rangle}{E_n-E_i+k'-i\varepsilon}
\right]\,.
\eeqa 
Obviously, the calculation of the resonance part is
more involved because of the summation over all possible
intermediate states $|I_n\rangle$ and energies $E_n$.

The low energy expansion of the polarizabilities has been discussed in
Ref.~\cite{ArW86}. For $k=0$ only the scalar E1-polarizability is
nonvanishing, i.e.
\beq
P_J(E1,E1)|_{k=0}=-\delta_{J0}\,\widehat I\,\sqrt{3}\,\frac{e^2Z^2}{M_A}\,,\label{low-energy}
\eeq
with $I$ as ground state spin, which corresponds to the Thomson amplitude.

\subsection{The scattering cross section}
Before turning to the LIT method, we will briefly review the
scattering cross section in terms of the polarizabilities. 
For unpolarized photon and target it is given by
\beq
\frac{d\sigma}{d\Omega}=\frac{k'}{k}\,\frac{c(\vec k,\vec p_i,k')}{2(2I_i+1)}
\sum_{\lambda,\lambda',M_i,M_f} |T^{fi}_{\lambda'\lambda,M_f,M_i}(\vec
k^{\,\prime},\vec k)|^2\,, 
\eeq
with a kinematic factor resulting from the final state phase space and
the incoming flux factor
\beq
c(\vec k,\vec p_i,k')=\frac{\omega+E_i
  -\omega'}{(\omega+E_i)|\frac{\vec k}{\omega}-\frac{\vec
    p_i}{E_i}|}\,. 
\eeq
In terms of the polarizabilities one finds for the cross section~\cite{ArG69,Wey88}
\beqa
\frac{d\sigma}{d\Omega}&=&\frac{k'}{k}\,\frac{ c(\vec k,\vec p_i,k')}{2I_i+1}
\sum_{L',L,K',K,J}\,\,\sum_{\nu',\nu,\bar\nu',\bar\nu}
P_{if,J}(M^{\nu'}L',M^{\nu}L) P_{if,J}^*(M^{\bar\nu'}K', M^{\bar\nu}K)
g_J^{\nu' L' \nu L \bar\nu' K' \bar\nu K}(\theta)\,,
\eeqa
where the angular functions are given by
\beqa
g_J^{\nu' L'  \nu L  \bar\nu' K' \bar\nu K}(\theta)&=&
\frac{(-)^J}{2}(2J+1)(-)^{L+K+\nu'+\bar \nu'}\sum_{j}(2j+1)
(1+(-)^{L+K+j+\nu+\bar\nu})
(1+(-)^{L'+K'+j+\nu'+\bar\nu'})\nonumber\\
&&\times
\left(\begin{matrix}
L'&K'& j \cr 1 & -1&0 \cr
\end{matrix}\right)
\left(\begin{matrix}
L&K& j \cr 1 & -1 &0 \cr
\end{matrix}\right)
\left\{\begin{matrix}
L&K& j \cr K' & L'&J \cr
\end{matrix}\right\} P_j(\cos\theta)\,,
\eeqa
with $ P_j(\cos\theta)$ as Legendre polynomials. For pure $E1$
transitions one obtains 
\beq
\frac{d\sigma (E1)}{d\Omega}=\frac{k'}{k}\,\frac{c(\vec k,\vec p_i,k')}{(2I_i+1)}
\sum_{J}|P_{if,J}(E1,E1)|^2\,g_J^{E1}(\theta)\,,
\label{xsect}
\eeq
where in an abbreviated notation
\beqa
g_0^{E1}(\theta)&=&\frac{1}{6}\,(1+\cos^2\theta)\,,\\
g_1^{E1}(\theta)&=&\frac{1}{4}\,(2+\sin^2\theta)\,,\\
g_2^{E1}(\theta)&=&\frac{1}{12}\,(13+\cos^2\theta)\,.
\eeqa

\subsection{Application of the Lorentz integral transform method}

A convenient method for the evaluation of the polarizabilities is
provided by the Lorentz Integral Transform (LIT)~\cite{EfL07} 
as applied to exclusive reactions. 
For this purpose we separate in Eq.~(\ref{compact-res}) the
intermediate propagator from the reduced matrix element by writing 
\beq
G(k+i\epsilon)=\int_{E_0}^\infty dE
\frac{\delta(H-E)}{E-E_i-k-i\epsilon} \,,
\eeq
where $E_0$ denotes the ground state energy, 
and introduce for the separated reduced matrix element as a convenient
abbreviation a quantity which henceforth will be called the
polarizability strength function  
\beq
F_{(\nu'L',\nu L)J}^{I_fI_i}(k',k,E)=\frac{(-)^{J+I_f+I_i}}{\widehat J}\langle
I_f E_f||\left[M^{\nu',L'}(k')\,\times\delta(H-E)\,
M^{\nu,L}(k)\right]^J||I_i E_i\rangle\,.\label{strengthfunction-a}
\eeq
One should note that in general the strength function is off energy
shell, i.e.\ $E\neq E_i+k$.
Evaluating the reduced matrix element, one obtains
\beq
F_{(\nu'L',\nu L)J}^{I_fI_i}(k',k,E)=\sum_{I_n}\rho(I_n,E)
\left\{\begin{matrix}
L&L'& J \cr I_f &I_i& I_n \cr 
\end{matrix}\right\}\,
\langle I_f E_f||M^{\nu',L'}(k')||I_n,E\rangle
\langle I_n,E||M^{\nu,L}(k)||I_i E_i\rangle\,,\label{strengthfunction}
\eeq
with $\rho(I,E)$ as density of states for a given energy $E$
and angular momentum $I$. 
In terms of the strength functions the polarizability becomes
\beqa
P_{if,J}^{\rm res}(M^{\nu'}L',M^\nu L,k',k)&=&2\pi(-)^{L+I_f+I_i}\hat L\hat L'
\nonumber\\&& 
 \times\int_{E_0}^\infty dE\Big[\frac{F_{(\nu'L',\nu L)J}^{I_fI_i}(k',k,E)}
{E-E_i-k-i\varepsilon}
+(-)^{L+L'+J}
\frac{F_{(\nu L,\nu' L')J}^{I_fI_i}(k,k',E)}
{E-E_i+k'-i\varepsilon}\Big]\,.
\eeqa
One can separate the real and imaginary parts of the propagator
according to 
\beqa \label{P-Re}
(P_{if,J}^{\rm res}(M^{\nu'}L',M^\nu L,k',k))_{\rm Re}&=&2\pi(-)^{L+I_f+I_i}\hat L\hat L'
\nonumber\\&& \times 
{\cal P}\int_{E_0}^\infty dE\Big[\frac{F_{(\nu'L',\nu L)J}^{I_fI_i}(k',k,E)}
{E-E_i-k}
+(-)^{L+L'+J}
\frac{F_{(\nu L,\nu' L')J}^{I_fI_i}(k,k',E)}
{E-E_i+k'}\Big]
\,,
\eeqa
where ${\cal P}$ stands for the principal value of the integral, and
\beqa
(P_{if,J}^{\rm res}(M^{\nu'}L',M^\nu L,k',k))_{\rm Im} &=&2\pi^2(-)^{L+I_f+I_i}\hat L\hat L'
\nonumber\\&& 
\times
\Big[F_{(\nu'L',\nu L)J}^{I_fI_i}(k',k,E_i+k)
+(-)^{L+L'+J}F_{(\nu L,\nu' L')J}^{I_fI_i}(k,k',E_i-k')
\Big]\,.
\label{P-Im}
\eeqa
where the second term contributes only if $E_i>E_0$, i.e.\ if the
initial state is an excited state. The subscripts ``Re'' and ``Im''
indicate the contributions of the real and imaginary parts
of the propagator, respectively. For elastic scattering the strength
function is real and then Eqs.~(\ref{P-Re}) and (\ref{P-Im}) represent
the real and imaginary parts, respectively, of the 
polarizability. 

The strength functions are the principal quantities which are determined by
the Lorentz Integral Transform method. 
Thus the main task is the evaluation of the strength function
$F_{(\nu'L',\nu L)J}^{I_fI_i}(k',k,E)$. To this end we first consider the
following partial strength function for a fixed intermediate total
angular momentum state $|I_n M_n\rangle$ as defined by
\beq
F_{\nu'L',\nu L}^{I_fI_i;I_n}(k',k,E)=\rho(I_n,E)
\langle I_f E_f||M^{\nu',L'}(k')||I_n,E\rangle
\langle I_n,E||M^{\nu,L}(k)||I_i E_i\rangle\,.\label{partialstrengthfunction}
\eeq
In terms of these partial strength functions the polarizability
strength is given by
\beq
F_{(\nu'L',\nu L)J}^{I_fI_i}(k',k,E)=\sum_{I_n}
\left\{\begin{matrix}
L&L'& J \cr I_f &I_i& I_n \cr 
\end{matrix}\right\}\,
F_{\nu'L',\nu L}^{I_fI_i;I_n}(k',k,E)\,. \label{strengthfunction-b}
\eeq
The partial strength can be expressed in terms of states of good total
angular momentum $I_n$ which are generated by the action of a multipole
operator $M^{\nu L}$ on a state $\psi^{I}$ with good angular momentum
$I$ according to
\beq
|(M^{\nu,L}(k)\times\psi^{I})I_nM_n\rangle=
\Big[M^{\nu,L}(k)\times|I\rangle\Big]^{I_n}_{M_n}\,.
\eeq
Namely, using
\beqa
\langle IM|(M^{\nu,L}(k)\times\psi^{I_i})I_nM_n\rangle&=&\sum_{MM_i}
(-)^{I_i-L-M_n}\hat I_n 
\left(\begin{matrix}
L&I_i& I_n \cr M &M_i& -M_n \cr 
\end{matrix}\right)\,\langle IM|M^{\nu L}_M|I_iM_i\rangle\nonumber\\
&=&\delta_{II_n}\delta_{MM_n}\frac{(-)^{I_i-I_n-L}}{\hat I_n}
\langle I_n||M^{\nu L}||I_i\rangle\,,
\eeqa
one finds
\beqa
F_{\nu'L',\nu
  L}^{I_f,I_i;I_{n} }(k',k,E)&=&(-)^{I_n-I_i+L-L'+\nu'}\sum_{M_n}
\langle (M^{\nu',L'}(k)\times\psi^{I_f})I_n M_n
|\delta(H-E) |(M^{\nu,L}(k)\times\psi^{I_i})I_n M_n\rangle
\,.
\label{partial-str-0}
\eeqa
Then we perform a Lorentz integral transform with a
complex argument 
$\sigma=\sigma_R+i\sigma_I$ 
\beqa \label{LIT}
L_{\nu' L',\nu L}^{I_f,I_i;I_n}(k',k,\sigma)&=&
\int_{E_0}^\infty dE\,\frac{F_{\nu'L',\nu
    L}^{I_f,I_i;I_n}(k',k,E)}{(E-\sigma)(E-\sigma^*)}\,.
\eeqa
Inserting the explicit form for the strength function of
Eq.~(\ref{partial-str-0}) and integrating over the $\delta$-function, one
finds consecutively 
\beqa
L_{\nu'L',\nu L}^{I_f,I_i;I_{n}}(k',k,\sigma)&=&
(-)^{I_n-I_i+L-L'+\nu'}\nonumber\\
&&\times\sum_{M_n}
\langle (M^{\nu',L'}(k)\times\psi^{I_f})I_n M_n
|(H-\sigma)^{-1}(H-\sigma^*) ^{-1}|(M^{\nu,L}(k)\times\psi^{I_i})I_n M_n\rangle
\nonumber\\  
&=&(-)^{I_n-I_i+L-L'+\nu'}
\rho(I_{n},\sigma)\sum_{M_n}\langle \widetilde\psi_{I_f;I_{n}M_n}^{\nu',L'}(k',\sigma)|
\widetilde\psi_{I_i;I_{n}M_n}^{\nu,L}(k,\sigma)\rangle\,,
\eeqa
where $\rho(I_{n},\sigma)$ takes into account the possibility, that for the given $I_n$
and $\sigma$ several Lorentz states may exist. 
Here, the Lorentz state of good total angular momentum $I_n$ and
projection $M_n$ obeys the equation
\beq
(H-\sigma^*)
|\widetilde\psi_{I_i;I_{n}M_n}^{\nu,L}(k,\sigma)\rangle= 
|(M^{\nu,L}(k)\times\psi^{I_i})I_nM_n\rangle\,.\label{lorentz-eq}
\eeq

After inversion of the Lorentz transform, one obtains the desired
polarizability strength function from Eq.~(\ref{strengthfunction-b})
which then serves for the evaluation of the appropriate generalized
polarizability. 

\section{The Deuteron Case}
For the calculation of deuteron elastic photon scattering  the
$\gamma$-deuteron c.m.-system is chosen, where one has $k=k'$. 
In order to demonstrate the LIT method it suffices to consider the
at low energies dominant $E1$ transitions with the $E1$ operator in
the long wave length approximation (Siegert form)
\beq
E^1_M=i[H,D^1_M]\,,\mbox{ where }D^{1}_M = {\frac
  {\sqrt{\alpha}}{3\sqrt{2}} }\, r \,Y_{1M}(\Omega) 
\eeq
is independent of $k$. Here $\alpha$ denotes the fine structure
constant, and $(r,\Omega)$ 
represents the relative neutron-proton coordinate. 
Furthermore, $H$ denotes the intrinsic two-nucleon hamiltonian
containing the nucleon-nucleon interaction. The small c.m.\
contribution to the Hamiltonian is neglected for simplicity. 
Due to the dipole 
approximation only the polarizabilities $P_J(E1,k)$ ($J=0,1,2$, in an
abbreviated notation) contribute.  Instead of the corresponding
partial $E1$-strength function
\beqa
F_{E1, E1}^{11;j}(E) &=&(-)^{j-1}\sum_m
\langle ([D^{1},H]\times\psi^1_d)jm|\,\delta(H-E)\,|([D^{1},H]
\times\psi^1_d)jm\rangle\nonumber\\ 
&=&(-)^{j-1} (E-E_0)^2 \sum_m\langle (D^{1}\times\psi^1_d)jm|\,\delta(H-E)\,|(D^{1}
\times\psi^1_d)jm\rangle\,,
\eeqa
where $E_0$ denotes the ground state energy,
we will consider the reduced partial strength function
\beqa
\widetilde F_{E1, E1}^{11;j} (E) &=&
\frac{F_{E1, E1}^{11,j} (E)}{(E-E_0)^2} \nonumber\\
&=&(-)^{j-1}\sum_m\langle (D^{1}\times\psi^1_d)jm|\,\delta(H-E)\,|(D^{1}
\times\psi^1_d)jm\rangle\,.
\eeqa
One should note that $\widetilde F_{E1, E1}^{11;j} $ is independent of
$k$. The associated Lorentz state obeys as LIT equation (see
Eq.~(\ref{lorentz-eq})) 
\beq \label{D-L-state}
(H-\sigma^*)|\widetilde\psi_{jm}(\sigma)\rangle=
|(D^{1}\times \psi^1_d)jm\rangle\,,
\eeq
which is independent of $k$. The Lorentz state can be expanded into
partial waves according to the orbital angular momentum $l$
\beq
\langle r,\Omega |\widetilde\psi_{jm}(\sigma)\rangle=
\frac {\sqrt{\alpha}}{r \sqrt{4\pi}}\sum_{l=|j-1|}^{j+1}
\Phi_{jl}(\sigma,r) 
\langle \Omega|(l1)j \,m\rangle \,,
\eeq
where $|(l1)j \,m\rangle $ represents the spin-angular state of
orbital angular momentum $l$ coupled with total spin-one to a
total angular momentum $j$, and the matrix element refers to spin and
angular degrees of freedom only. The state
$\widetilde\psi_{jm}(\sigma)$ generates the LIT $\widetilde
L^{11;j}_{E1,E1}$ of the reduced strength $\widetilde
F^{11;j}_{E1,E1}$. 

Inserting the expansion of $\widetilde\psi_{jm}$ into
Eq.~(\ref{D-L-state}) and projecting onto a state $|(l1)j \,m\rangle$, 
one finds a set of radial differential equations
\beq
\label{LITeq_rdl}
\left[ -{\frac {\hbar^2}{M}} \left( {\frac {d^2}{dr^2}} - {\frac {l(l+1)}{r^2}}
 \right) - \sigma^*\right] \Phi_{jl}(\sigma,r)+ \sum_{l'}V_{jl,jl'}
\Phi_{jl'}(\sigma,r) = {\frac {1}{3 \sqrt{2}} }\, r f_{jl}(r) 
\eeq
with
\beq
f_{jl}(r) = \delta_{l1} \, u(r) + \,\,(-)^{j+1} \, 3 \, \sqrt{5} \, {\hat l} 
\left(\begin{matrix}
2&1&l \cr 0 &0&0 \cr
\end{matrix}\right)
\left\{\begin{matrix}
2&1&1 \cr j &1&l \cr
\end{matrix}\right\} \, w(r) \,,
\eeq
where $u(r)$ and $w(r)$ are the deuteron radial $s$- and $d$-wave
functions. Two of them are uncoupled $(^3P_0: l=1,j=0;$ $^3P_1:
l=1,j=1)$ and one is coupled $(^3P_2-\,^3F_2: l_1=1,l_2=3,j=2)$. 
Equation (\ref{LITeq_rdl}) is very similar to a radial Schr\"odinger
equation, but with a complex energy $\sigma$ and an additional source
term on the right-hand side. 

For convenience we introduce the following three reduced LITs
\beq
\label{LIT_j}
L_j (\sigma):=(-)^{j-1} \frac{4\pi}{2j+1} \widetilde L^{11;j}_{E1,E1}(\sigma) = %
\frac{4\pi}{2j+1}\sum_m\langle \widetilde\psi_{jm}(\sigma)|
\widetilde\psi_{jm}(\sigma)\rangle=
\alpha\sum_{l} 
\int_0^\infty |\Phi_{jl}(\sigma,r)|^2 dr \,,
\eeq
and corresponding reduced strength $F_j(E)=\widetilde
F^{11;j}_{E1,E1}(E)/(2j+1)$. Then the polarizability strength function
becomes
\beq
F_{E1, E1}^{11;j}(E) =\frac{(E-E_0)^2}{4\pi}\sum_j (-)^{j+1}(2j+1)
\left\{\begin{matrix}
1&1&J \cr 1&1&j \cr
\end{matrix}\right\}F_j(E)\,,\label{FE1E1-J}
\eeq
and, according to Eq.~(\ref{P-Im}), the $E1$ polarizabilities are given by
\beqa
(P_J^{\rm res}(E1,k))_{Im} &=&-6\pi^2\,F_{E1, E1}^{11;j}(k+E_0)\nonumber\\
&=&\frac{3}{2}\pi \,k^2\sum_j (-)^j(2j+1)
\left\{\begin{matrix}
1&1&J \cr 1&1&j \cr
\end{matrix}\right\}F_j(k+E_0)\,,\label{Im-PJ}\\
(P_J^{\rm res}(E1,k))_{Re}&=&\frac{3}{2}\sum_j (-)^j(2j+1)
\left\{\begin{matrix}
1&1&J \cr 1&1&j \cr
\end{matrix}\right\}
{\cal P}
\int dE (E-E_0)^2 F_j(E) \Big( \frac{1}{E-E_0-k}+\frac{(-)^J}{E-E_0+k}\Big)\nonumber\\
&=&\frac{1}{\pi} {\cal P} \int dk' \,
(P_J^{\rm res}(E1,k'))_{Im}\Big( \frac{1}{k'-k}+\frac{(-)^J}{k'+k}\Big)\,. \label{Re-PJ}
\eeqa
The latter expression for the real part corresponds to the dispersion
theoretic approach of Ref.~\cite{WeA83}. It is a consequence of the
fact, that in this special case of taking the $E1$-operator in the low
energy limit the polarizability strength $F_{E1, E1}^{11;j}$ becomes independent of the
photon momentum $k$. However, in general this is not true.

\section{ Results and Discussion}

For the numerical solution of the radial equation (\ref{LITeq_rdl})
for the radial Lorentz states $\Phi_{jl}(\sigma,r) $ we have chosen 
the 
Argonne potential AV18~\cite{AV18} as interaction model. 
\begin{figure} [ht]
\centerline{\includegraphics*[scale=.4]{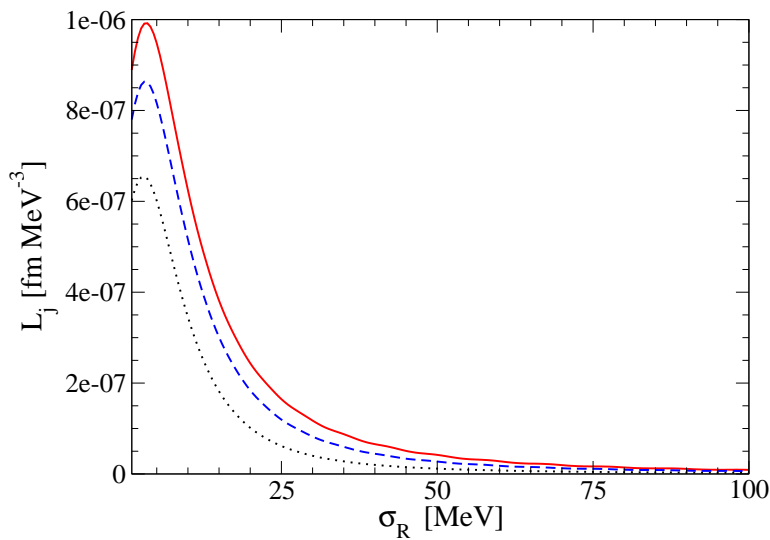}}
\caption{(Color online) Lorentz integral transforms $L_j(\sigma)$ with $\sigma_I=5$
  MeV: $j=0$ (dotted), $j=1$ (solid), $j=2$ (dashed).} 
\label{LIT_j_100}
\end{figure}
The resulting three $L_j(\sigma)$ are shown in
Fig.~\ref{LIT_j_100} for a constant $\sigma_I=5$~MeV as function of
$\sigma_R$ up  to  100 MeV.
The figure shows that the three LITs have quite a similar
behavior. All three exhibit a pronounced peak in the low $\sigma_R$
region, only the peak heights are slightly different. For the
principal value integral in (\ref{P-Re}) also high-energy
contributions could play an important role, therefore we illustrate in
Fig.~\ref{LIT_j_1000} the transforms in an extended
$\sigma_R$ range.   
 \begin{figure} [ht]
\centerline{\includegraphics*[scale=.4]{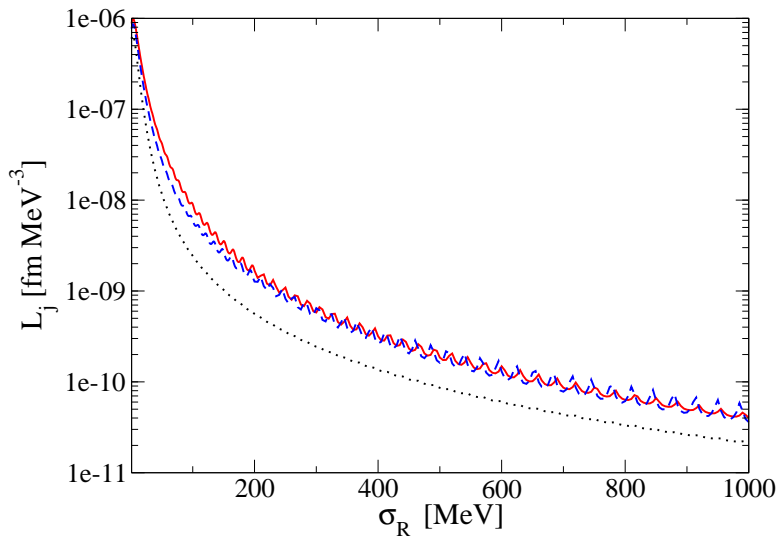}}
\caption{(Color online) As in Fig.~\ref{LIT_j_100}, but for an
  extended range of $\sigma_R$.}
\label{LIT_j_1000}
\end{figure}
One finds that the behavior at high $\sigma_R$ is approximately
described by $\sigma_R^{-2}$. 
It shows that at large $\sigma_R$ the transforms are dominated by
low-energy contributions. In fact, for the extreme case of a
$\delta$-strength, i.e. $F(E)=F_0 \,\delta(E-E_0)$, one obtains
$L(\sigma)=F_0/((\sigma_R-E_0)^2+\sigma_I^2)$.  

One also notices in Fig.~\ref{LIT_j_1000} the onset of oscillations
at higher $\sigma_R$, which are more pronounced for
$L_1$ and $L_2$. The origin of these oscillations lies in the
relatively small 
value of 5 MeV for $\sigma_I$, which makes a high-precision solution
of (\ref{LITeq_rdl})  with increasing $\sigma_R$ more and more
difficult. Such a small $\sigma_I$ value, however, is advantegeous in
the low $\sigma_R$ region, where (i) it does not lead to numerical
problems and (ii) prominent structures of a small width could be
present, e.g, a low-energy peak. The small $\sigma_I$ value will then
allow one to resolve such details and nonetheless will not lead to
problems for the reconstruction of the high-energy strength, since
structures with a small width are not present at higher energies. One
could also completely avoid the oscillations at higher $\sigma_R$ by
choosing a transform with different $\sigma_I$ values for low- and
high-energy regions, as has been done in~\cite{ELOT10}. 

In order to obtain the strength function $F_{j}(E)$ one has
to invert the integral
transforms $L_{j}$ defined in (\ref{LIT_j}). 
Details about the inversion of the LIT are found
in~\cite{EfL07,AnL05}, and further, more general inversion aspects are 
discussed in~\cite{BELO10}. Accordingly, we use expansions of the
calculated LITs, $L_j(\sigma_R,\sigma_I=5$MeV), in a set of
basis functions $\widetilde \chi_n^{(1)}$ ($n=1,2,...,N$), where the
expansion coefficients are determined by a best fit. Here we take, as
in \cite{EfL07}, the set (for fixed $\sigma_I$)
\beq
\widetilde \chi_n^{(1)}(\sigma_R) = \int_0^\infty dE {\frac {\chi_n^{(1)}(E)}
{(E-\sigma_R)^2 + \sigma_I^2} }
\eeq
with
\beq \label{basis}
\chi_n^{(1)} (E) =  E^{\alpha_1} exp(-{\frac {\alpha_2 E}{n \beta} }) \,,
\eeq
where $\alpha_i$ and $\beta$ are nonlinear parameters. 
\begin{figure}
\centerline{\includegraphics*[scale=.4]{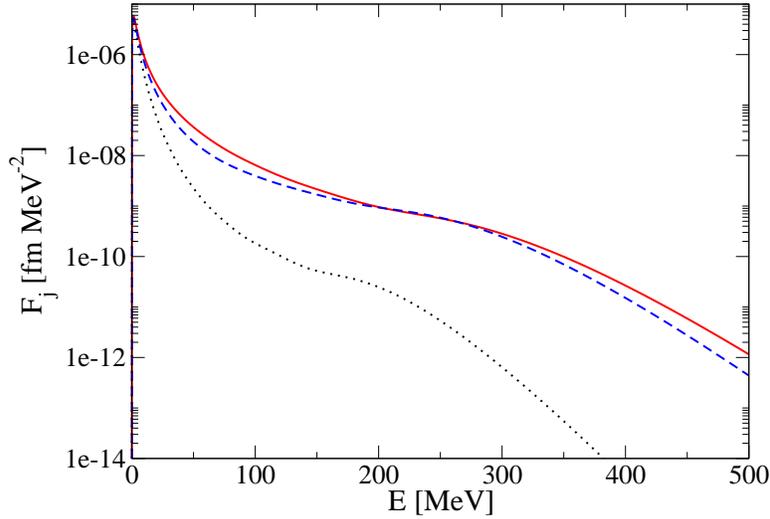}}
\caption{(Color online) Strength function $F_{j}(E)$ for $j=0$
  (dotted), $j=1$ (solid), $j=2$ (dashed).} 
\label{F_j}
\end{figure}
The  inversion results, $F_j(E)$, are shown in Fig.~\ref{F_j}. 
One observes pronounced
low-energy peaks with a strong subsequent fall-off which becomes
weaker at somewhat higher energies. The fall-off becomes stronger
again at even higher energies, namely beyond about 200 MeV (300 MeV)
in case of $F_0$ ($F_{1/2}$). As a matter of fact, the inversion
results are already not very good somewhat below those energies, since
there we do not find, as it would be necessary, a stability for the
inversion solutions for a limited range of the number of basis
functions $N$. The origin of this problematic high-energy behavior
lies in the choice of basis
functions. By construction they all have an exponential
high-energy fall-off and thus are not suitable to describe a function with
a different high-energy behavior, since $N$ cannot be 
increased arbitrarily because of the numerical accuracy of the
transform to be fitted. 

\begin{figure}
\centerline{\includegraphics*[scale=.4]{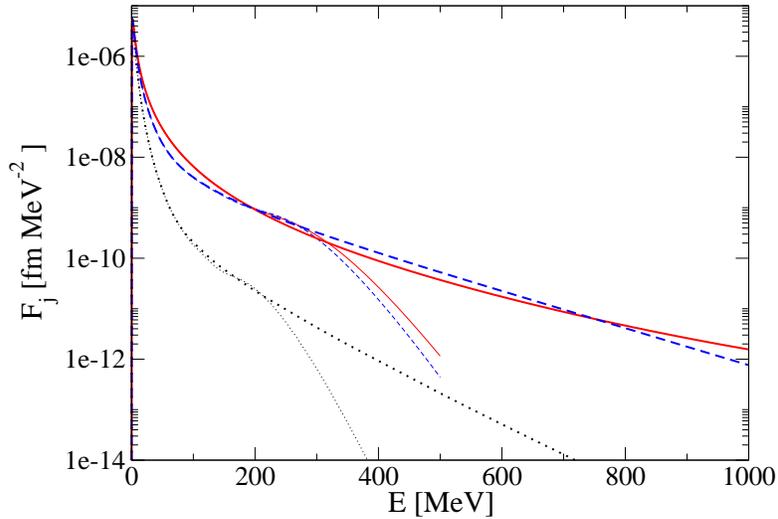}}
\caption{(Color online) Strength functions $F_j$ from inversions with first (thin
  lines) and second basis set (thick lines): $j=0$ (dotted), $j=1$
  (solid), $j=2$ (dashed).} 
 \label{F2_j}
\end{figure}

As already mentioned, for the principal value integral in (\ref{P-Re})
also high-energy contributions of the strength function could
matter. Therefore it is better to search for a basis set which does
not have the shortcomings of the set in 
(\ref{basis}), but which is more appropriate to describe the strength
function over a much larger energy range, even if it could lead to a
somewhat less precise inversion at lower energies. For this purpose we
introduce an alternative set without an exponential fall-off: 
\beq \label{basis2}
\chi^{(2)}_n =  E^{\alpha_n(E)} 
\eeq
with
\beq 
\alpha_n(E) = (n+ {\frac {1}{2}}) + \left(\beta - (n + {\frac
    {1}{2}})\right) \left({\frac {E}{E_{\rm asy}} }\right)^\gamma \,, 
\eeq
where $\beta < 0$, $\gamma$, and $E_{\rm asy}$ are nonlinear
parameters. In the present case we have calculated the LIT up to 1000
MeV and thus we set $E_{\rm asy}=1000$ MeV, i.e. all basis functions
have an asymptotic fall-off with $E^\beta$ at $E=1000$ MeV. 
The inversions with basis set $\{\chi^{(2)}\}$ are shown in
Fig.~\ref{F2_j}. At lower energies one finds a similar picture as in
Fig.~\ref{F_j}, the high-energy behavior, however, is quite
different. One further notes that a constant asymptotic high-energy
fall-off is already reached at energies considerably lower than 
$E_{\rm asy}=1000$ MeV showing that this set is a very good choice.
\begin{figure}
\centerline{\includegraphics*[scale=.4]{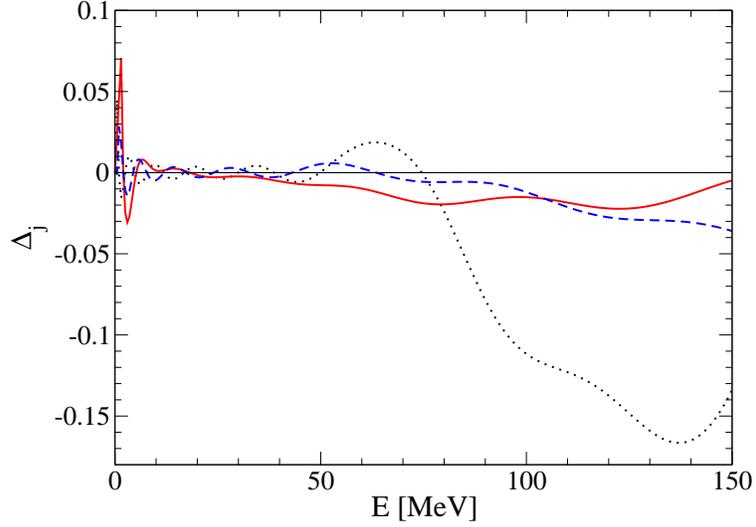}}
\caption{(Color online) Relative differences $\Delta_j(E)$ of the strength
  functions shown in Fig.~\ref{F2_j}. Notation of curves as in
  Fig.~\ref{F2_j}.} 
\label{compare}
\end{figure}

Figure~\ref{F2_j} also shows that at lower energies the inversion
results do not depend significantly on the choice of the basis set. On
the other hand the precision of the agreement is difficult to judge on
a logarithmic scale. Therefore, we show in Fig.~\ref{compare} the
various inversions at lower energies in a more detailed formn, i.e. as
relative differences
\beq
\Delta_j(E)=(F^{((1)}_j(E) -  F^{(2)}_j(E))/F^{(2)}_j(E)\,,
\eeq
where  $F^{(1)}_j$ and $F^{(2)}_j$ correspond to the inversions with
the first and the second basis set, respectively. At very
low energies one finds the largest differences for $F_1$, namely about
7\%. This relatively large difference is due to the steep rise of the
strength function right above threshold. Here we would like to
mention that, as suggested before, our first basis set leads to a better fit of the
calculated LITs in the low $\sigma_R$ region. Furthermore, 
Fig.~\ref{compare} shows that up to 50 MeV further differences remain 
very small, in fact less than about 1\%. 
For $F_1$ and $F_2$ the picture does not change very much for higher
energies, whereas in case of $F_0$ the difference increases quite
substantially beyond 80 MeV. This can be interpreted as a precursor effect for
the strong fall-off beyond 200 MeV which leads to a somewhat
oscillating inversion result already at considerably lower energies. 
In order to have the best description in the whole energy range
we combine the inversion results of the two basis sets by
taking $F^{(1)}_j(E)$ for $E \le E_1$ and $F^{(2)}_j(E)$ for $E \ge E_1$
with $E_1$ equal to 30, 18, and 33 MeV for j=0, 1, 2, respectively.

\begin{figure}
\centerline{\includegraphics*[scale=.6]{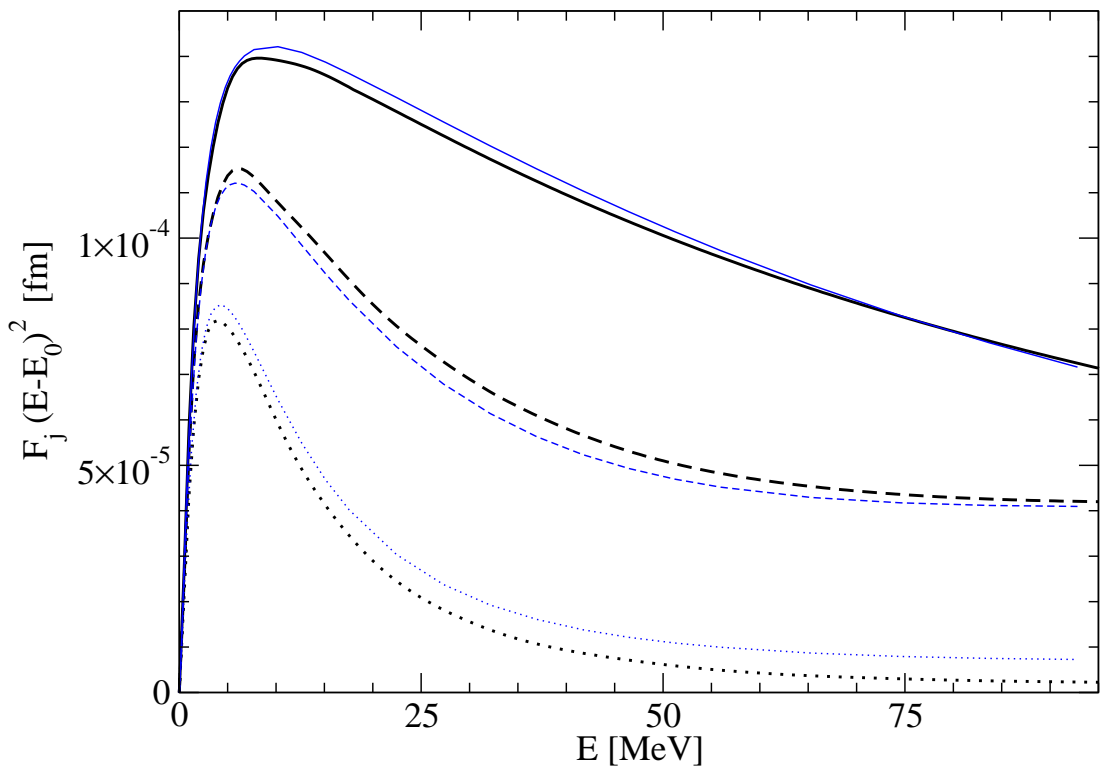}}
\caption{(Color online) Strengths functions $F_{j}(E)$ weighted with $(E-E_0)^2$
  for the Argonne V18  potential
  in standard approach with complete $E1$-operator (thin lines) and LIT
  result with low energy $E1$-operator (thick lines): $j=0$ (dotted),
  $j=1$ (solid), $j=2$ (dashed).} 
\label{comp-Fj}
\end{figure}

The strength functions $F_j$ can also be obtained from a standard
calculation of deuteron photodisintegration. According to
Ref.~\cite{WeA83} one has 
\beqa
\mathrm{Im}\, P_J(E1,E1)&=&4\pi k\sum_j(-)^j
\left\{\begin{matrix}
1&1& J \cr 1 &1&j \cr
\end{matrix}\right\}
\sum_{\mu ls} |E^1(\mu,jls)|^2\,,\label{polariz_a}
\eeqa
where the $E1$-matrix elements are defined by the multipole expansion 
of the total unpolarized photo absorption cross
section~\cite{Par64,ArS91} 
\beq
\sigma_0=\frac{(4\pi)^2}{3}\,
\sum_{L\mu,jls} \frac{1}{2L+1}(|E^L(\mu,jls)|^2+|M^L(\mu,jls)|^2)\,.
\eeq
Here, the quantum numbers $(\mu ls)$ classify the various components
of a two-body scattering solution with total angular momentum
$j$~\cite{Par64,ArS91}. 
Comparison with Eq.~(\ref{Im-PJ}) yields the relation
\beq
F_j(k+E_0)=\frac{8}{3 k(2j+1)}\sum_{\mu ls} |E^1(\mu,jls)(k)|^2\,.
\eeq
A comparison of the standard calculation with the LIT approach is
shown in Fig.~\ref{comp-Fj}. The agreement is quite satisfactory in
view of the fact, that in the standard calculation the complete
$E1$-operator is used and not its low energy form as in the
present LIT approach.

\begin{figure}
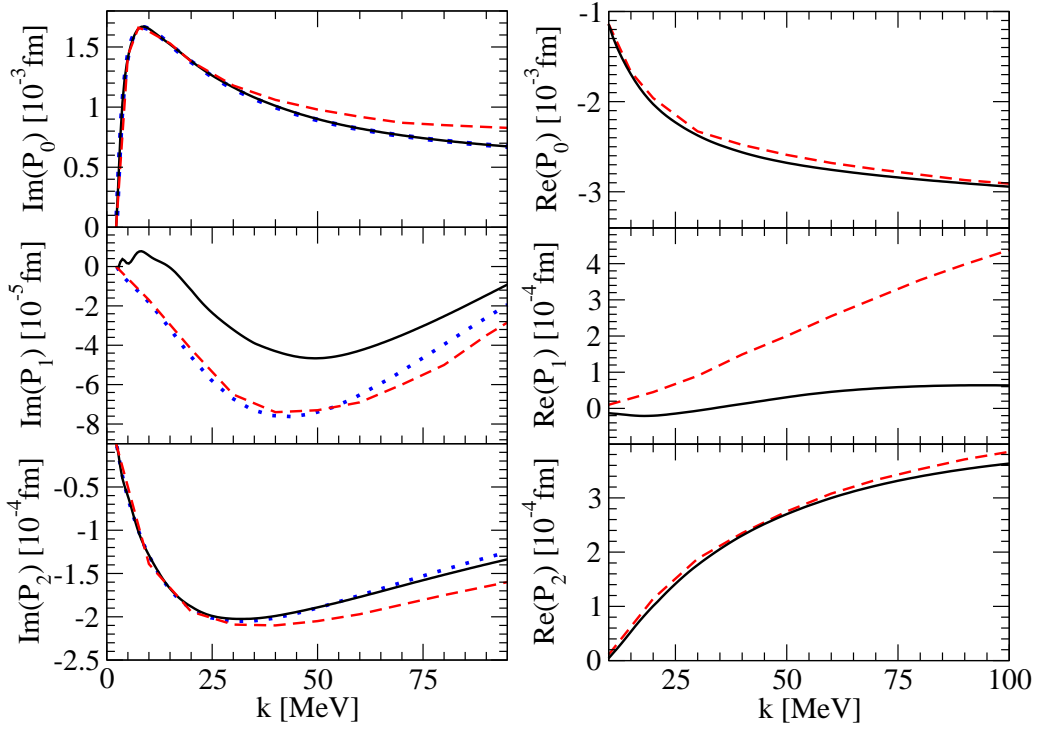

\centerline{\includegraphics*[scale=.5]{photon_sct_Fig9.eps}
{\includegraphics*[scale=.5]{photon_sct_Fig10.eps}}} 
\caption{(Color online) Imaginary (left panels) and real (right panels) parts of the
  polarizabilities $P_J^{\rm res}(E1,E1,k)$ (solid) in comparison to
  those of \cite{WeA83} (dashed): $J=0$ (top panels), $J=1$ (middle
  panels), $J=2$ (bottom panels). Note that the real parts are normalized to
  zero at $k=0$ (see text). For the imaginary parts also the results with
  from the standard calculation are shown (dotted)}
\label{ImP_j}
\end{figure}

Having obtained the reduced strength $F_j(E)$, one can then determine
with the help of Eqs.~(\ref{Im-PJ}) and (\ref{Re-PJ}) the imaginary
and real parts of the polarizabilities $P_J^{\rm res}(E1,E1,k)$. In 
Fig.~\ref{ImP_j}  we compare our results for the polarizabilities to
those of \cite{WeA83}. One should note that the real parts are
normalized to zero at $k=0$ for $J=0$ and 2, which is not necessary
for $J=1$ since one has $P_J^{\rm res}(E1,E1,k=0)=0$. 
This normalization takes into account
implicitly the neglected contributions of the two-photon
amplitude at $k=0$, which are needed in order to comply with the low energy
theorem of Eq.~(\ref{low-energy}) for $J=2$, whereas for the
scalar polarizability we have to add the classical Thomson limit to
the normalized resonance part according to
\beq
{\rm Re}\,(P_0(E1,E1,k)) = {\rm Re}\,(P_0^{\rm res}(E1,E1,k)) - {\frac {3e^2}{m_d}} \,,
\eeq
where $m_d$ is the deuteron mass. This procedure is justified
according to the discussion in \cite{Wey88} (see Fig.~3 of
\cite{Wey88}), where it is shown that further contributions to the
two-photon amplitude beyond the low energy limit 
are negligibly small for $k \le 40$ MeV and remain quite
small up to about 60 MeV. Thus, for calculations below 60 MeV, it
seems to be quite safe to simply add the Thomson term to the
normalized resonance E1E1 scalar polarizability. 

With respect to the comparison in Fig.~\ref{ImP_j}, we should mention
that the polarizabilities of \cite{WeA83} are calculated in the Breit
system. The relation between the photon momentum $k$ of the Breit to
that of the cm system depends on the 
scattering angle. Only in forward direction they turn out to be
exactly the same. In addition it should be mentioned that in
\cite{WeA83} 
a more complete calculation has been made where the full nuclear
one-body current and also pion exchange currents have been taken into
account. 
Furthermore, in
\cite{WeA83} a different NN potential has been used. Despite these
differences, one finds a rather good agreement between the two
calculations for the scalar and tensor polarizabilities. Larger
differences are found for the vector polarizability, for which
real and imaginary parts from \cite{WeA83} are somewhat larger
than our results. In Fig.~9 also the imaginary parts of the $P_J$
evaluated by the standard method are shown. For $J=0$ and 2 they
are very similar to our results, actually the agreement is even better 
than one would expect from the results of Fig.~8. On the other hand, for $J=1$ the 
standard result agrees quite well with that of
Ref.~\cite{WeA83}. However, one should keep in mind, that the vector
polarizability is about an order of magnitude smaller than the scalar
one due to large cancellations between the $F_j$. Thus the small
differences in the $F_j$ according to Fig.~\ref{comp-Fj} result in
considerably larger differences in $P_1$. 

\begin{figure}
\centerline{\includegraphics*[scale=.4]{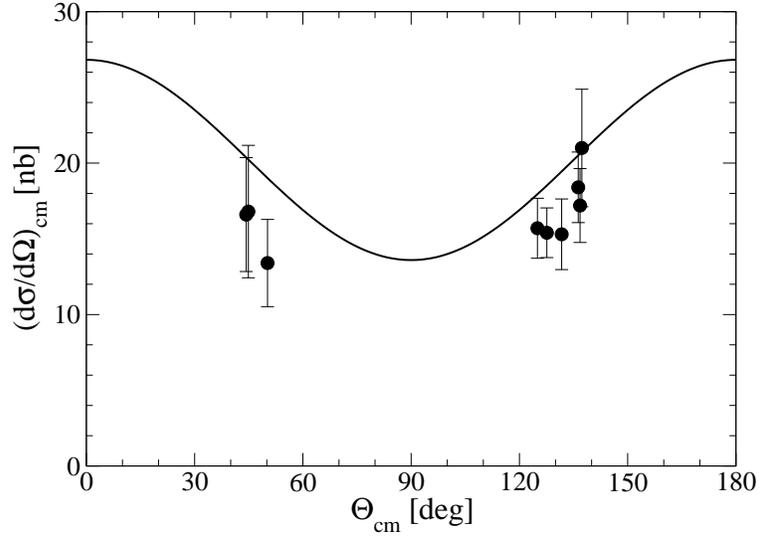}}
\caption{Comparison of the differential scattering cross section in
  the unretarded dipole approximation with experimental data from
  \cite{Lun03} ($k_{\rm lab}=55$ MeV).} 
\label{wq}
\end{figure}

With the calculated polarizabilities we can determine the cross
section for pure E1 transitions given in (\ref{xsect}). 
In Fig.~\ref{wq} we the show the resulting cross section in comparison
to experimental data at $k_{\rm lab}=55$ MeV. One sees that our
theoretical result overestimates slightly the experimental cross
section, although within the experimental errors. 
Since data exist only for a very limited angular range one cannot draw
any conclusion whether the E1E1 cross section reproduces the correct
angular shape. However, from other
theoretical calculations~\cite{WeA83,KaM99,LeL01,HiG05,HiG10}
mentioned above, it is known that additional polarizabilities
$P_J^{\rm res}(M^{\nu'}L',M^\nu L,k)$ from neglected other multipoles,
e.g.\ $M1$ and $E2$, play a considerably less important role at lower energies.  
Furthermore, in Refs.~\cite{KaM99,LeL01,HiG05,HiG10}, it is pointed out that
also the internal nucleon 
polarizabilities lead to non-negligible contributions. In fact, one aim
of present-day Compton scattering experiments on light nuclei is the
determination of the nucleon static electric and magnetic dipole
polarizabilities, in particular the ones of the neutron.

We summarize our work as follows. After a brief 
overview over the theory of photon scattering, we
have described as the central issue of our work the application of
the LIT method to the calculation of the photon scattering amplitude. This
method has the great 
advantage over conventional methods, since the resonance
amplitude can be evaluated in a very efficient way, in which wave functions
of the continuum spectrum of a particle system have not to
be calculated explicitly. Because of this reason, and in contrast to
conventional state-of-the-art methods, the LIT approach can
be applied also to systems with more than three particles. 

As a first application and test case, we have considered elastic
photon scattering off the deuteron at low energies. For this case we
have taken into account only E1 transitions in the unretarded dipole
approximation. In order to apply the LIT method 
one has to invert the calculated integral transforms of the so-called
strength functions. This inversion is discussed in great detail. In
particular the high-energy behavior of the strength functions is
studied with great care. We have compared the resulting three E1E1
polarizabilities with results from conventional calculations. The comparison has
shown the reliability of the obtained results. At last we have
confronted the E1E1 cross section with experimental data at a photon
energy of 55 MeV. In view of the approximations of the present work
the agreement is quite satisfactory. This is certainly a strong
encouragement to consider in future work photon scattering off more
complex light nuclei.

\end{document}